\documentclass{article}
\usepackage{amssymb,amsmath,amsthm,graphicx,epsfig,latexsym}

\font\german=eufm10 at 10pt
\def\Buchstabe#1{{\hbox{\german #1}}}

\def\EA{\Buchstabe{A}}      
\def\Z2{\mathbb{Z}_2}
\def\f{\phi}            

\newtheorem{theorem}{Theorem}

\newtheorem{lemma}{Lemma}

\newtheorem{claim}{Claim}

\newcommand{\ket}[1]{|#1\rangle}
\newcommand{\bra}[1]{\langle#1|}
\newcommand{\braket}[2]{\langle#1|#2\rangle}
\newcommand{\m}[1]{\overline{#1}}

\begin{document}


\vfill


\vfill
\vfill

\begin{center}
   \baselineskip=16pt
   \begin{LARGE}
      \textsl{The Kochen-Specker Theorem Revisited\\*[0.6em]
        in Quantum Measure Theory}
   \end{LARGE}
   \vskip 2cm
      Fay Dowker and Yousef Ghazi-Tabatabai
   \vskip .6cm
   \begin{small}
      \textit{Blackett Laboratory, Imperial College,\\
        London, SW7 2AZ, U.K.}
        \end{small}
\end{center}

\vskip 1cm

\begin{small}
\begin{center}
   \textbf{Abstract}
\end{center}
The Kochen-Specker Theorem is widely interpreted to imply that non-contextual
hidden variable theories that agree with the predictions of Copenhagen
quantum mechanics are impossible. The import of the theorem for
a novel observer independent interpretation of quantum mechanics,
due to Sorkin, is investigated.

\end{small}
\vskip 0.5cm

%
%
%
%

\section{Introduction}

Though quantum mechanics is extremely successful in making
experimental predictions it has proved unusually difficult to
understand or interpret, to the extent that the interpretation
of quantum mechanics has become a field of research in itself.
While an instrumentalist approach evades these problems in
the context of open systems with external observers to hand,
in the
application of quantum mechanics to closed systems, as called for
by the search for a quantum theory of gravity, a notion of
objective reality becomes all the more necessary.
The spacetime nature of reality as revealed by Special and
General Relativity also puts pressure on the text-book
version of quantum theory in which a vector in a Hilbert space
plays a central role.

As argued by Hartle \cite{Hartle:1991bb} the development of the Sum
Over Histories (SOH) by Dirac \cite{Dirac:1933} and Feynman marked
the beginnings of a new framework for quantum mechanics in which a
more observer independent, and fully spacetime view of reality could
be pursued. Indeed Feynman's original paper is entitled ``Space-time
approach to non-relativistic quantum mechanics'' \cite{Feynman:1948}
and in his Nobel Lecture he described how ``[he] was becoming used
to a physical point of view different from the more customary point
of view. [...] The behavior of nature is determined by saying her
whole spacetime path has a certain character.'' \cite{Feynman:1965}

In a SOH framework, the structure of a quantum theory is
more akin to a classical stochastic theory such as Brownian
motion than any classical Hamiltonian theory,
the kinematics consisting of a space of
histories or ``formal trajectories'' of the system and the dynamics
being a quantal generalisation of a probability measure
on that space.
Due to the presence of interference between histories,
however, the quantal measure cannot be interpreted as
a probability distribution and the challenge has been
to furnish the SOH
with an interpretation that would make it an
observer independent theory consistent with experimental
observation.

To date, the most widely studied attempt to interpret the SOH is the
``consistent histories'' or ``decoherent histories'' approach and in
its SOH formulation (as opposed to its ``projection operator'' form)
has been championed primarily by Hartle \cite{Hartle:1989,Hartle:1992as}. In
this approach, the interpretational difficulty that interference
causes is dealt with by imposing the rule that only partitions of
the space of histories (coarse grainings) such that there is no residual
interference between elements of the partition are allowed to be
considered. The quantal measure, when restricted to the elements of
such a partition, behaves like a classical probability measure and
this is the basis for the decoherent histories interpretation.

Sorkin has proposed an alternative approach to finding an
interpretation of the SOH \cite{Sorkin:2006wq,Sorkin:2007}
starting from an analysis
of the interpretational structure of classical stochastic theories.
He identified three basic structures in such a classical theory: a
Boolean algebra, $\EA$, of ``questions'' that can be asked about the
system (or ``propositions'', ``predicates'' or ``events''), a space
of ``answers'' (or ``truth values'') which classically is $\Z2 =
\{0,1\}= \{false, true\} = \{no, yes\}$, and the space of allowed
``answering maps'' $\phi: \EA \rightarrow \Z2$ which classically is
the space of non-zero homomorphisms. Each answering map, or
{\it co-event}, $\phi$,
corresponds to a possible reality and its being a homomorphism is
equivalent to the use of Boolean (ordinary) logic to reason about
classical reality.

To generalise this structure to the quantal case,
Sorkin proposes to keep both the ``event algebra'' $\EA$ and the
truth values, $\Z2$ as they are but relax the condition that
the answering map/co-event $\phi$ must be a homomorphism. Thus,
the general framework can be described as one that
embraces ``Anhomomorphic Logic''.

The freedom to incorporate Anhomomorphic Logic
enables the identification of co-events $\phi$
as potential objective
realities in a quantum theory.  The framework is
 therefore challenged to provide an account of the
standard counter-examples to the classical realist
approach, in particular the celebrated Kochen-Specker theorem
\cite{Kochen:1967}. This paper will go some way to
showing that this is indeed possible
in a specific proposed scheme
which we will call, after Sorkin, the
Multiplicative Scheme.

First we outline Quantum Measure Theory
and describe the Multiplicative Scheme.
We then review the
Peres version of the Kochen-Specker theorem, and
recast it in the language of
quantum measure theory.
We show that although there is no homomorphic
co-event for the Peres-Kochen-Specker event algebra there
is a multiplicative
co-event.
We then show that this quantum measure theory
version of the Peres setup can be
concretely realised as a
Peres-Kochen-Specker gedankenexperiment in terms of the
spacetime paths of a spin 1 particle
though a sequence of beam splitters and recombiners.
We sketch an account of this gedankenexperiment
in the Multiplicative
Scheme.

\section{Quantum Measure Theory}\label{section:quantum measure theory}

The kinematics of a classical stochastic theory, such as a random walk, is specified by a space, $\Omega$, of possible
histories or formal trajectories. For simplicity we will assume here and in the quantal case that all spaces are finite. Then the event algebra, $\EA$, referred to in the introduction is the Boolean algebra of all subsets of $\Omega$.

The dynamics of a classical stochastic theory is
encoded in a measure, $|\cdot |$ such that
\begin{align}
|\cdot |&:\ \ \EA \rightarrow\mathbb{R}\nonumber\\
|A|&\geq0 \ \  \forall A \in \EA \ \ \text{(Positivity)}\nonumber\\
|A\sqcup B| &=|A| + |B|\ \
\forall A,B \in \EA\,, \ \text{where} \ A, B \
\text{disjoint}\ \ \text{(Kolmogorov Sum rule)}\nonumber \\
|\Omega| &= 1\ \ \text{(Normalisation)}\label{eq:measure normalisation} \,.
\end{align}

Elements of $\EA$ will be referred to as
events and each event corresponds to a question that can be
asked about the system.

One can think of a possible reality as giving a set of true/false answers to every question that can be asked, in other words it is a map, $\phi$, from $\EA$ to $\Z2$, a {\it co-event} (so called because it takes events to numbers). The set of all such maps will be denoted $\EA^*$. Now a classical stochastic theory distinguishes exactly one of the fine grained histories (say $\gamma\in\Omega$) as the `real history'.  A question $A\in\EA$ is also a subset of $\Omega$, and is `true' if and only if it contains the real history $\gamma$. We can then express this state of affairs as a co-event $\gamma^*\in\EA^*$ mapping to $1$ (true) exactly those elements of $\EA$ that contain the real history $\gamma$:
\begin{align*}
*&:\ \   \Omega \rightarrow \EA^*\\
*&:\ \  \gamma \mapsto \gamma^* \,,
\end{align*}
where
\begin{align*}
\gamma^*(A) = &\, \left\{\begin{array}{cc} 1 & \text {if}\ \gamma \in A \\
                                     0 & \text {otherwise} \, \end{array}\right.
\end{align*}

The ``mind flip'' from formal trajectory, $\gamma$, to co-event, $\gamma^*$, as representing reality turns out to be
fruitful in moving on to the quantum case. We will refer to co-events such as $\gamma^*$,
where $\gamma$ is an element of $\Omega$, as {\it classical} co-events.

$\EA$ is a boolean algebra with the operations symmetric 
difference as addition  and intersection as multiplication. 
$\mathbb{Z}_2$ is also a Boolean algebra and 
the classical co-events are precisely the homomorphisms 
from $\EA$ to $\mathbb{Z}_2$.
\begin{lemma}Given a finite sample space $\Omega$ and an associated event algebra $\EA=2^\Omega$
\begin{equation}
\{\gamma^* | \gamma\in\Omega\}=Hom(\EA,\mathbb{Z}_{2})
\end{equation}
where the zero map is excluded from $Hom(\EA,\mathbb{Z}_{2})$.\footnote{
Throughout this paper we disregard the zero co-event.}
\end{lemma}

Understanding the role of the measure $|\cdot|$ is controversial, even in a classical stochastic theory
where it is tantamount to understanding the interpretation of probability. Sorkin, following Geroch \cite{Geroch:1984}
among others, has adopted the view that the predictive, scientific content of a stochastic theory
(classical or quantal) is exhausted by the requirement of ``preclusion'' whereby an event of zero
measure will be valued ``false'' in reality. In other words, allowed co-events, $\gamma^*$, are ``preclusive''\footnote{In order for a stochastic theory to make successful
predictions in the case of finitely many trials of a random process
(a coin toss, say), this preclusion rule must be widened to say that
an event of very small measure is also precluded.
In this form -- and especially when it is understood as
containing {\it all} the predictive content of
a stochastic theory -- the rule is known as Cournot's principle
and was the dominant interpretation of probability held
during the first half of the 20th century \cite{Shafer:2005}.
Quantum Measure Theory will have to embrace
Cournot's Principle in order successfully to reproduce the
probabilistic predictions of Copenhagen quantum mechanics but
for the purposes of this paper, the strictly zero measure rule will suffice.}:

\begin{equation}\label{preclusion}
|A| = 0 \Rightarrow \gamma^*(A) = 0
\end{equation}

 Thus if an event $A\in\EA$ is precluded by the measure, $|A|=0$, then any co-eve
nt that is an `admissible' description of the system is of the form $\gamma^*$ 
and must obey $\gamma^*(A)=0$.
This is equivalent to the condition that the real history, $\gamma$,
cannot be an element of $A$. This is straightforward and 
uncontroversial but it is exactly this condition that will cause problems with a classical interpretation of quantum mechanics, as we shall see.

The structure of a quantal measure theory \cite{Sorkin:2006wq, Sorkin:2007} is the same as the classical one in many respects and we will use the same notation: a space, $\Omega$ of formal trajectories and the corresponding Boolean
algebra of events, $\EA$, give the kinematics of the theory. The dynamics is encoded in a measure which is again a
non-negative, normalised, real function, $|\cdot|$, on the event algebra. However, the Kolmogorov Sum Rule is replaced by its quantal analogue:
\begin{equation}\label{eq:sumrule}
|A\sqcup B\sqcup C| = |A\sqcup B| + |B\sqcup C| + |A\sqcup C| - |A|
- |B| - |C| \ \  \forall A, B, C \in \EA, \ \text{disjoint}\,
\end{equation}
which is just one of a whole hierarchy of conditions that define higher level ``generalised measure theories'' \cite{Sorkin:1994dt}. This replacement leads to important differences with the classical theory, in particular, in contrast to a probability measure, due to interference a measure zero set can now contain sets that have positive measure. More generally, $A\subset B$ no longer implies $|A|\leq|B|$.

An alternative (more or less equivalent \cite{Sorkin:1994dt})
set of axioms for the dynamical
content of a quantal
theory is based on the so-called ``decoherence functional'',
$D: \EA \times \EA \rightarrow \mathbb{C}$ which satisfies the
conditions
 \cite{Hartle:1989,Hartle:1992as}:

\noindent (i) Hermiticity: $D(X;Y) = D(Y;X)^*$ ,\  $\forall X, Y\in \EA$;

\noindent (ii) Additivity: $D(X\sqcup Y; Z) = D(X;Z) + D(Y;Z)$ ,\
$\forall X, Y, Z \in \EA$ with $X$ and $Y$ disjoint;

\noindent (iii) Positivity: $D(X;X)\ge0$ ,\  $\forall X \in\EA$;

\noindent (iv) Normalisation: $D(\Omega ;\Omega)=1$  .

\noindent The quantal measure is given by
a decoherence functional via
\begin{equation}
|X| = D(X ; X)\, .
\end{equation}
and it satisfies the condition in equation \ref{eq:sumrule}.
All known quantum theories have their dynamics encoded as
a decoherence functional, from which the quantal measure is
defined as above. To see how the
decoherence functional is defined for ordinary non-relativistic
particle quantum mechanics see for example \cite{Hartle:1992as,
Hartle:2006nx}.

Reality is to be identified with a co-event, $\phi$, an element of
$\EA^*$ as before but in the quantal case,
there are obstructions to demanding that the
co-event be a homomorphism.
The celebrated Kochen-Specker theorem
is one such obstruction and we will explore this explicitly in the next
section.

\subsection{The Multiplicative Scheme}\label{sec:mult scheme}

A homomorphic co-event respects both the sum (``linearity'') and
product (``multiplicativity") structure of the algebra $\EA$:
\begin{align*}
\f(A+B) = &\,\f(A) + \f(B) \ \ \text{and}\\
\f(AB)  = &\,\f(A)\f(B) \ \ \forall A, B \in \EA \,,
\end{align*}
where
$A+B = (A\setminus B) \cup (B\setminus A)$ is symmetric difference
and $AB = A \cap B$ (recall that
$A$ and $B$ are identified with subsets of $\Omega$).
The Multiplicative Scheme \cite{Sorkin:2007}
for Anhomormorphic Logic can be motivated from several directions.
Algebraically,
it can be defined by requiring that real co-events preserve
multiplicativity,
\begin{equation*}
\phi(AB) = \phi(A)\phi(B) \ \forall A,B \in \EA\;,
\end{equation*}
while the linearity condition is dropped. The two trivial co-events
$\phi = 1$ and $\phi = 0$ are excluded by fiat in the scheme.

\begin{lemma}If $\phi$ is multiplicative then the set of
events $\phi^{-1}(1)$,
\begin{equation*}
\phi^{-1}(1) = \{ A \in \EA | \phi(A) = 1\}\, ,
\end{equation*}
is a filter.
\end{lemma}
\begin{proof}
Let $\phi$ be a multiplicative co-event and let
$A \in \phi^{-1}(1)$. If $A \subset B$ then
$AB = A$ and therefore $1 = \phi(A) = \phi(AB) = \phi(A)\phi(B) = \phi(B)$.
Which means that $B \in \phi^{-1}(1)$. Further, if $A,~B\in\phi^{-1}(1)$
then $\phi(AB)=\phi(A)\phi(B)=1$, so $A\cap B\in\phi^{-1}(1)$,
thus $\phi^{-1}(1)$ is a filter.
\end{proof}

Note that this proof holds for infinite $\Omega$ also.
However, every finite filter is a principal filter, so if
$\Omega$ is finite (as we are assuming unless otherwise stated) then
$\phi^{-1}(1)$ is a principal filter. To see this first note that
because $\Omega$ and thus $\EA$ are finite, $\phi^{-1}(1)$ must
contain a minimal element (minimal under set inclusion) $\Phi$.
If $\Xi$ is also a minimal element of the filter then
$\Phi \cap \Xi$ is in the filter, is contained in both
$\Phi$ and $\Xi$ and therefore must be equal to both.
Therefore $\Xi=\Phi$ and $\Phi$ is the unique minimal element
and the principal element of
$\phi^{-1}(1)$. Note that $\Phi$ is not empty since otherwise
$\phi = 1$ which is excluded. In the terminology of reference \cite{Sorkin:2007},
$\Phi$ can also be called the support of the multiplicative
co-event $\phi$.

It is convenient to identify a multiplicative co-event $\phi$ with
its support, {\it i.e.} the principal element $\Phi$ of the associated filter.
We define a map
\begin{align*}
*&:\ \  \EA \rightarrow \EA^*\\
*&:\ \  A \mapsto A^* \,,
\end{align*}
where
\begin{align*}
A^*(B) = & \ \left\{\begin{array}{cc} 1 & {\text{if} ~A \subset B} \\
                                      0 & {\text{otherwise}}\end{array}\right.
\end{align*}
Note that, using this
notation, $\{\gamma\}^*=\gamma^*$.
The map $*$ is clearly 1-1 and its
image is the set of all
multiplicative co-events (for finite $\Omega$). The inverse of
$*$ is the map that takes a multiplicative co-event,
$\phi$, to the
principle element, $\Phi$, of the associated filter.

The natural order on filters
furnishes us with a simple notion
of minimality or {\it primitivity}.
We say that a preclusive multiplicative co-event
$\phi$ is minimal or primitive if the associated filter $\phi^{-1}(1)$
is maximal among the filters corresponding to preclusive multiplicative
co-events. Primitivity
can be thought of as a condition of ``maximal detail''
or ``finest graining'' consistent with preclusion (``Nature
abhors imprecision'') and we
adopt this as a further condition on real co-events in the
Scheme. This
allows us to {\it deduce}
classical logic when the measure happens to be
classical:

\begin{lemma}\label{lemma:multiplicative classical}
If $\phi$ is a primitive preclusive multiplicative co-event and the
measure $|\cdot|$ is classical, then $\phi$ is a homomorphism.
\end{lemma}
\begin{proof}
Consider $\Phi\in\EA$ such that $\phi = \Phi^*$. The
Kolmogorov sum rule implies that the union, $Z$, of measure zero sets in
$\EA$ is itself of measure zero, so the preclusivity of $\phi$
implies $\Phi\not\subset Z$. Thus $\exists \gamma \in \Phi\setminus
Z$ such that  $\gamma$ is not an element of any measure zero set.
This means that
$\gamma^*$ is preclusive, but then the minimality of $\phi$ means
that $\Phi = \{\gamma\}$ and so $\phi = \gamma^*$, a homomorphism.
\end{proof}

All non-zero multiplicative co-events are {\it unital},
in other words $\phi(\Omega) = 1$ when $\phi$ is multiplicative.
So a multiplicative co-event answers the question ``Does anything
at all happen?'' with ``Yes''!
A nonzero preclusive multiplicative co-event always exists, indeed
$\Omega^*$ is always a preclusive multiplicative co-event.
Moreover, if $A$ is an event with nonzero measure that is not
contained in any event with zero measure then $A^*$ is
preclusive.  Therefore
there exists at least one primitive preclusive multiplicative co-event
in which $A$ happens.

\section{The Kochen-Specker Theorem}

The Kochen-Specker (KS) theorem \cite{Kochen:1967} is often cited as
a key obstacle to a realist interpretation of quantum mechanics.
We will situate the KS theorem in the
framework of event algebras and co-events described
above.

We will use Peres' proof of the Kochen-Specker Theorem
\cite{Peres:1991} and rely on his 33-ray construction.

\begin{figure}
\begin{center}
\includegraphics[width=0.3\textwidth]{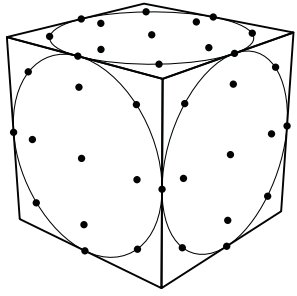}
\caption{\small{The $\pm 33$ directions are defined by the
lines joining the center of the cube to the
$\pm 6$ mid--points of the edges and the
$\pm3$ sets of 9 points of the $3 \times 3$ square arrays
shown inscribed in the incircles of its faces}} \label{Peres Set}
\end{center}
\end{figure}

Peres defines 33 rays in $\mathbb{R}^{3}$
from which 16 orthogonal bases can be formed.
Peres defines the rays as being the 33 for which the
squares of the direction cosines are one of the combinations:
\begin{equation*}
0+0+1=0+1/2+1/2=0+1/3+2/3=1/4+1/4+1/2 \,.
\end{equation*}
Figure \ref{Peres Set} (taken with permission from \cite{Conway:2006})
illustrates the points where the rays intersect a unit cube
centred on the origin.

To describe the rays, Peres employs a shorthand notation that we
will find useful, writing $\m{x} = -x$ we define
\begin{equation*}
\m{x}yz = \alpha\left(\begin{array}{c}
-x\\y\\z\end{array}\right)\in\mathbb{R}^{3}
\end{equation*}
 where $\m{xyz} = xyz$
as we are dealing with rays rather than vectors. We will call this
set of rays the Peres Set, $PS$, and sometimes it will be convenient
to refer to them as labelled from 1 to 33: $PS = \{u_i| i =
1,2,\dots 33\}$ (in some fixed but arbitrary order).

In what follows it will be important to understand the structure
\& the symmetries of $PS$. Examining the magnitude
of the angle between each ray and its nearest neighbour, we find
that the rays can be divided into four types:

\begin{table}[ht]
\begin{center}
\begin{tabular}{l}
Type I: $001,010,100$
\\Type II: $011,01\m{1},101,10\m{1},110,1\m{1}0$
\\Type III: $012,0\m{1}2,021,02\m{1},102,\m{1}02,201,20\m{1},120,\m{1}20,210,2\m{1}0$
\\Type
IV: $112,\m{1}12,1\m{1}2,\m{1}\m{1}2,121,12\m{1},\m{1}21,\m{1}2\m{1},211,21\m{1},2\m{1}1,2\m{1}\m{1}$
\end{tabular}
\label{table:Ray Types}
\caption{The Peres rays divide into 4 Types.}
\end{center}
\end{table}

With reference to figure \ref{Peres Set},
Type I corresponds to the midpoints of faces of the cube, Type II
to the midpoints of edges, Type III to the remaining points in the interior
of the incircles of the faces and Type IV
to the remaining points on the incircles of the faces.
It can be seen that each
symmetry of the projective cube induces a permutation of $PS$.
The symmetry group of the projective cube, $H$, is
of order 24 and is generated by rotations by
$\pi/2$ around co-ordinate axes and reflections in co-ordinate
planes. Since the ray types are defined in terms of angles, the induced
permutations on the elements of $PS$ will preserve the type of each
ray, {\it i.e.} the  permutation reduces to permutations on the four
subsets of same-type rays. By inspection, $H$ is transitive on each type:
this means that for any two rays $u$, $v$ of the same type $\exists$
$g\in H$ such that $v=g(u)$.

In the Peres version of the Kochen-Specker theorem, this
set of 33 rays is mapped in the obvious way into a set of rays
 in a 3 dimensional Hilbert space, where two rays are
orthogonal in the Peres set if and only if they are mapped
to orthogonal rays in Hilbert space. We will therefore talk about the
Peres rays both as rays in real 3-D space and in Hilbert space.
To each orthogonal
basis of rays,
$\{u_{i},u_{j},u_{k}\}$, in Hilbert space
can be associated an observable with
three distinct outcomes, each outcome corresponding to one of the
rays in the basis (the eigenvector of the outcome/eigenvalue).
If that observable is measured, one of the three outcomes will be
obtained and
the quantum state collapses onto exactly one of the three basis rays,
which corresponds to the outcome.
If it is assumed that
the result of the measurement
of the observable exists within the
system before or independent of the measurement being taken
-- a non-contextual hidden variable --
then we would conclude that
one of the basis rays corresponds to the actual value of the
observable being measured, and is labelled ``true'',
and the other two (using classical logic) are labelled ``false''. We
follow Peres in considering the the ``true'' ray to be coloured green
whereas
the other two are coloured red.

Assuming that an experimenter
can freely choose to measure any of the observables associated to the
16 orthogonal bases in $PS$, if we assume that the result of the
measurement actually done is encoded in the system beforehand,
then the results of all potential measurements must be encoded
in the system. Thus,
all of the rays in $PS$ will be coloured.
We call a map
$$\gamma:PS\rightarrow \{green,red\}$$
a colouring.
We call $\gamma$ a {\it consistent} colouring
if it colours exactly one ray green out of every basis in $PS$
(we assume henceforth that ``basis'' implies ``orthogonal
basis'') and does not colour any pair of orthogonal
rays both green. Note there are some orthogonal pairs in $PS$ that
are not contained in a basis in $PS$.

\begin{table}[ht]
\begin{center}
\begin{tabular}{ccc}
\hline
Basis & Basis Rays & Other Orthogonal Rays \\
\hline
$B_1$ & $\textbf{001}~100~010$ & $110~1\m{1}0$ \\
$B_2$ & $\textbf{101}~\m{1}01~\textit{010}$ & \\
$B_3$ & $\textbf{011}~0\m{1}1~\textit{100}$ & \\
$B_4$ & $\textbf{1$\m{\textbf{1}}$2}~\m{1}12~\textit{110}$ & $20\m{1}~021$ \\
$B_5$ & $\textbf{102}~\textit{20$\m{\textit{1}}$}~\textit{010}$ & $2\m{11}$ \\
$B_6$ & $\textbf{211}~\textit{0$\m{\textit{1}}$1}~\textit{2$\m{\textit{11}}$}$
& $\m{1}02$\\
$B_7$ & $\textbf{201}~\textit{010}~\textit{$\m{\textit{1}}$02}$ & $\m{11}2$\\
$B_8$ & $\textbf{112}~\textit{1$\m{\textit{1}}$0}~\textit{$\m{\textit{11}}$2}$
& $02\m{1}$\\
$B_9$ & $\textbf{012}~\textit{100}~\textit{02$\m{\textit{1}}$}$
& $\m{1}2\m{1}$\\
$B_{10}$ & $\textbf{121}~\textit{$\m{\textit{1}}$01}~\textit{$\m{\textit{1}}
$2$\m{\textit{1}}$}$ & $0\m{1}2$\\
$B_{11}$ & $\textit{100}~\textit{021}~\textit{0$\m{\textit{1}}$2}$ &
\end{tabular}
\caption{Peres' Proof of the Kochen-Specker Theorem}\label{table:Peres}
\end{center}
\end{table}

\begin{theorem}\label{thm:PKS}
 {\textbf{Peres-Kochen-Specker} \cite{Peres:1991}} \\There is no consistent
colouring of $PS$
\end{theorem}
\begin{proof}
The proof is based on table \ref{table:Peres}. First assume there is a
consistent colouring of $PS$. This colouring must be consistent on all
bases in $PS$, in particular it must be consistent on the four bases
$B_1,~B_2,~B_3,~B_4$. These bases intersect, and there are $24$
consistent colourings, of the $10$ rays in $\bigcup_iB_i$. It
is easy to see \cite{Peres:1991} that these 24 colourings are
related to each other by the 24
symmetries of the projective cube.
Consider a single fiducial colouring, $\gamma_P$ of the four bases,
$B_1,~B_2,~B_3,~B_4$, as defined in table \ref{table:Peres}, which colours
green the first ray in each basis except for $B_{11}$
(highlighted in bold in the table).
We then work down the table basis by basis starting at $B_5$
to try to extend $\gamma_P$ to a
consistent colouring of $PS$. For each basis $B_i$, $i>4$, in
turn we find
that two of the basis elements (italicised) have already been coloured
red, and so in each case the choice of basis ray to colour green is
forced by consistency. This continues until we reach basis $B_{11}$,
which by then has all three basis rays coloured red, meaning that $\gamma_P$
cannot be extended to a consistent colouring of the whole Peres Set.
Now, as noted above every consistent colouring
$\gamma_Q$ of $B_1,~B_2,~B_3,~B_4$ is related to $\gamma_P$ by a
symmetry, thus $\gamma_Q=g(\gamma_P)$ for some $g\in H$. Thus if
we were to attempt to extend $\gamma_Q$ to a consistent colouring
of the whole Peres Set we would find the same contradiction on
$g(B_{11})$. Thus no consistent colouring of $B_1,~B_2,~B_3,~B_4$
can be extended to $PS$, therefore there is no consistent colouring
of the Peres Set.
\end{proof}

\subsection{Event algebras and co-events}

Let us now restate the theorem in terms of event algebras and co-events. Let the space $\Omega$ be the set of all (green/red) colourings, $\gamma$ of the Peres set $PS$. The event algebra is the Boolean algebra of subsets of $\Omega$ as before. Given a subset $S$ of the Peres Set we write:
\begin{align}
R_S =& \{\gamma\in\Omega |~\gamma(u_i)= red ~\forall u_i\in S\} \\
G_S =& \{\gamma\in\Omega |~\gamma(u_i)= green ~\forall u_i\in S\} 
\end{align}

Now the Kochen-Specker-Peres result depends crucially upon disallowing, or precluding, non-consistent colourings. Thus in measure theory language, we wish to consider the sets corresponding to these failures in consistency to have measure zero (though we will not explicitly construct a measure until section \ref{sec:Stern-Gerlach}). These inconsistencies arise in two ways, an orthogonal pair of rays being coloured green or an orthogonal basis of rays being coloured red, thus we treat the following two types of set as if they are of measure zero:

\begin{align}\label{eq:kssets}
R_B && B~\text{an orthogonal basis} \\
G_P && P~\text{an orthogonal pair}
\end{align}

We will call these the Peres-Kochen-Specker (PKS) events, or PKS sets. Note that the Peres Set includes orthogonal pairs that are not subsets of orthogonal bases.

Now (as already implicitly assumed in the proof of Theorem \ref{thm:PKS}) every symmetry $g\in H$ of the projective cube induces an action on $\Omega$ and thus on $\EA$, via its action on $PS$. We will also denote this action by $g\in H$, so that $g\cdot\gamma(u_i)=\gamma(g\cdot u_i)$ and $g(A) = \{g(\gamma) |\gamma\in A\}$. Crucially, note that the PKS sets are permuted by the symmetries in $H$, via $g(G_P)=G_{g(P)}$ and $g(R_B)=R_{g(B)}$.

In the language of measures and 
co-events, the Peres-Kochen-Specker result can be stated:
\begin{lemma}\label{failure of classical scheme}
Let $|\cdot |$ be a measure on the space $\Omega$ of colourings of $PS$ that is zero valued on the PKS sets. Then there is no preclusive classical co-event for this system.
\end{lemma}
\begin{proof}
By Theorem \ref{thm:PKS}, every element of $\Omega$ is inconsistent on at least one
basis or pair and therefore lies in at least one PKS set. Consider  $\gamma^*$,
a classical 
co-event. $\gamma$ lies in at least one PKS set and therefore $\gamma^*$ is not 
preclusive. \end{proof}

\section{Kochen-Specker in the Multiplicative Scheme}

Can multiplicative co-events succeed where the classical scheme fails in providing a satisfactory account of the PKS 
system? In particular, can we avoid non-existence results such as lemma 
\ref{failure of classical scheme}? Trivially, yes, since $\Omega^*$ is 
always a preclusive multiplicative co-event regardless of the measure, 
because we have {\it a priori} imposed $|\Omega |\neq 0$ (equation 1). 
However this is neither interesting nor useful, since the only event mapped to $1$ by $\Omega^*$ is $\Omega$ itself. In other words the only question answered in the affirmative is, 
``Does anything occur?''. 

$\Omega^*$ is too coarse grained to be useful and, indeed, the axioms of the 
Scheme are that only 
the finest grained
or {\it {primitive}} preclusive co-events are potential realities. 
So we need to examine co-events
that are minimal among the multiplicative
co-events mapping all the PKS sets and their disjoint
unions to zero.

Every
multiplicative co-event
is a principal filter, in particular every
multiplicative co-event is of the form $A^*$ for some $A\in\EA$ where
\begin{equation*}
A^*(B)=\left\{\begin{array}{cc}1 & A\subset B \\ 0 &
otherwise\end{array}\right.
\end{equation*}
$\forall~B\in \EA$. Hence a multiplicative co-event, $A^*$,
that values the PKS sets and their disjoint unions zero,
has as its support $A$ that is not
contained in any PKS set or any disjoint union of PKS sets.

The Peres-Kochen-Specker theorem
says that every $\gamma\in\Omega$ is an element
of at least one PKS set.
Some elements of $\Omega$, however, lie in exactly one PKS set and
so are good places to start building a
primitive co-event. If we can find two
elements $\gamma,\gamma' \in \Omega$ that each lie in exactly
one PKS set, different for each and not disjoint, then the set containing $\gamma$ and $\gamma'$
will be not contained in any PKS set or disjoint union
of PKS sets.

Looking at the proof of
Theorem \ref{thm:PKS}, 
we see there is a unique was to extend $\gamma_P$ from the 
four bases $B_1$, $B_2$, $B_3$, $B_4$ to the whole of $PS$ 
so that it is consistent on all bases 
and pairs of rays
except basis $B_{11}$.
We will henceforth refer to this
extension as the Peres colouring and use the same
notation $\gamma_P$ for it.
The Peres colouring lies in exactly one PKS set, $R_{B_{11}}$ 
(see equation \ref{eq:kssets}). 
$\gamma_P$ is given in table \ref{table:mult}.

We can obtain another such colouring $\gamma_P'$ by acting on $\gamma_P$
by one of the symmetries of the projective cube: let us choose
the reflection that exchanges the $x$ and $y$ axis. The colouring
$\gamma_P'$ is then obtained from $\gamma_P$ by permuting the
rays by the action of swapping the first and second labels
(in the Peres notation
we have used where a ray is labelled by the components, e.g. $001$).
$\gamma_P'$ is also given in table \ref{table:mult}.
By symmetry, $\gamma_P'$ is also contained in exactly one PKS set,
$R_{B_7}$.

The PKS sets $R_{B_7}$ and $R_{B_{11}}$ are not
disjoint because they share, for example, the colouring which
is red on all rays.

Therefore, the multiplicative co-event
$\phi_M \equiv \{\gamma_P, \gamma_P'\}^*$  will value all
PKS sets and disjoint unions of PKS sets zero and it is minimal
among multiplicative co-events that do so because neither of
the atomic co-events will do so.

\subsection{Events that can happen}

Suppose an event, $X$, has non-zero measure and is not
contained in an event of zero measure. Then that event
is the support of a preclusive, multiplicative
co-event, $X^*$ that values $X$ as $1$.
So we are guaranteed that a primitive preclusive co-event
that values $X$ as $1$ exists: that event could happen.

Let us assume that there is a quantum measure
on $\Omega$ the set of colourings of the Peres rays in
which the only sets of measure zero are the PKS sets
and their disjoint unions. Then for any Peres ray
$u_k$, the two events ``$u_k$ is coloured green'' and
``$u_k$ is coloured red'' both have non-zero measure and
are not contained in any PKS set and the following
constructions produce concrete examples of
primitive preclusive co-events in which these events happen.

Let $u_k$ be one of the
Peres rays and let $G_k \in \EA$
be the event that $u_k$ is coloured green and
$R_k \in \EA$ be the event that
$u_k$ is coloured red.
such that $\phi^r_{Mk}(R_k) = 1$ and $\phi^g_{Mk}(G_k) = 1$.

We start with $\phi_{M}$ (as defined above)
which maps the PKS sets to zero and is
minimal among the multiplicative events that do so.
Every symmetry of the cube induces
an action on the space of multiplicative co-events by
simultaneously acting on all colourings
in the support of a co-event.
Moreover, since the PKS sets are mapped to each other under
the symmetries of the cube, any co-event which values
the PKS sets zero will be mapped, under a symmetry, to
one which also values the PKS sets zero.

Table \ref{table:mult} shows the valuation of $\phi_{M}$ on the colourings
of all the rays in PS. We note that among the rays, $u_i$ for which
$\phi_{M}(G_i) =1$
 is at least one of each type (see list \ref{table:Ray Types}).
So given a ray $u_k$, we can find one, $u_{k'}$ of the same type
for which $\phi_M(G_{k'}) =1$. There is a symmetry of the cube
which sends $u_{k'}$ to $u_k$.
Acting with that symmetry on $\phi_M$ produces a co-event,
call it $\phi_{Mk}^g$, with the required properties.

Similarly, among the rays $u_i$ for which $\phi_M(R_i) =1$
are at least one of each type so we can find one, $u_{k''}$,
which of the same type as $u_k$. Acting on $\phi_M$ with a
symmetry that takes $u_{k''}$ to $u_k$ produces
a co-event, $\phi_{Mk}^r$, with the required properties.

\begin{table}[ht]
\begin{center}
\begin{tabular}{|c|ccc|cc|}
\hline
 $u_{i}$& & $\gamma_P$& $\gamma_P'$& $\phi_{M}(G_{i})$& $\phi_{M}(R_{i})$ \\
\hline
$001$& & g& g &1 &0\\
$010$& & r& r &0 &1\\
$100$& & r& r &0 &1\\
$011$& & g& g &1 &0\\
$01\m{1}$& & r& r &0 &1\\
$101$& & g& g &1 &0\\
$10\m{1}$& & r& r &0 &1\\
$110$& & r& r &0 &1\\
$1\m{1}0$& & r& r &0 &1\\
$012$& & g& g &1 &0\\
$0\m{1}2$& & r& r &0 &1\\
$021$& & r& g &0 &0\\
$02\m{1}$& & r& r &0 &1\\
$102$& & g& g &1 &0\\
$\m{1}02$& & r& r &0 &1\\
$201$& & g& r &0 &0\\
$20\m{1}$& & r& r &0 &1\\
$120$& & r& r &0 &1\\
$\m{1}20$& & r& r &0 &1\\
$210$& & r& r &0 &1\\
$2\m{1}0$& & r& r &0 &1\\
$112$& & g& g &1 &1\\
$\m{1}12$& & r& g &0 &0\\
$1\m{1}2$& & g& r &0 &0\\
$\m{11}2$& & r& r &0 &1\\
$121$& & g& g &1 &0\\
$12\m{1}$& & r& r &0 &1\\
$\m{1}21$& & r& r &0 &1\\
$\m{1}2\m{1}$& & r& r &0 &1\\
$211$& & g& g &1 &0\\
$21\m{1}$& & r& r &0 &1\\
$2\m{1}1$& & r& r &0 &1\\
$2\m{11}$& & r& r &0 &1\\
\hline
\end{tabular}
\caption{A minimal unital multiplicative co-event sending the PKS sets to zero}\label{table:mult}
\end{center}
\end{table}

\subsection{A closer look at $\phi_M$}\label{section:phi_M}

Looking at the table, we see that
for most of the rays we have $\phi(G_i) + \phi(R_i) = 1$ which is the
familiar situation in classical logic, for example if it is false that
$u_i$ is green then it is true that it is red.
However, there are some rays, $021, 201, \m{1}12$ and $1\m{1}2$,
for which both $\phi(G_i)=0$ and $\phi(R_i)=0$.
Note that there are no rays for which $\phi(G_i)=1$ and $\phi(R_i)=1$.
This is not a coincidence: $\phi(X)\phi(1+X) = \phi(0) = 0$
($\phi(0) = 1$ would imply $\phi =1$ since every event contains
the empty event) and so
an event and its complement cannot both be valued true. An event
and its complement {\it can} both be valued false as we see here
and indeed this is the epitome of the nonclassical logic that the
scheme tolerates.

This violation of classical rules of inference may
at the first encounter seem unpalatable. However, it is the way in
which a genuine coarse graining is expressed at the level of
logical inference. One way to think of the multiplicative co-event is
that reality corresponds to the support of the co-event and only
{\it common} properties of all the formal trajectories in the
support are real properties: if the support is contained in an
event then that event happens. In this way, both an event and its
complement -- {\it e.g.} ray $u_i$ is green and ray $u_i$ is red
-- can be valued false if the support of the co-event intersects them both.

The step in accepting the logic of multiplicative
co-events may not be so great, therefore, but we have to
deal with the question of whether such a co-event as $\phi_M$
represents a contradiction with experimental facts. Can not
we check to see whether $u_i$ is red and whether it is green
and if we always find it is one or the other (and never neither)
then wouldn't that contradict the proposal that
$\phi_M$ is a possible reality? We will address this
question in the context of the concrete experimental setup
of the next section.

\section{Expressing the Peres-Kochen-Specker setup in terms of Spacetime
Paths}\label{sec:Stern-Gerlach}

Thus far our analysis has been completely
abstract and we now seek to embed our mathematical
Peres-Kochen-Specker system into a more physical,
even perhaps experimentally realisable,
framework.
To this end we  imagine a
sequence of Stern-Gerlach apparatus to translate each colouring
in $\Omega$ into
a spacetime history. We will construct a quantum measure
on the corresponding
event algebra in which the translations of the Kochen-Specker sets
have measure zero, though further `accidental' measure
zero sets are introduced as will be discussed.

\subsection{The Stern-Gerlach Apparatus}

The Stern-Gerlach apparatus
allows spin
to be expressed in terms of
paths in spacetime.
In the original experiment (1922) a beam of spin $1/2$ silver atoms was
sent through an inhomogeneous magnetic field, splitting the beam
into two components according to the spin of the particles.

\begin{figure}
\begin{center}
\includegraphics[width=0.7\textwidth]{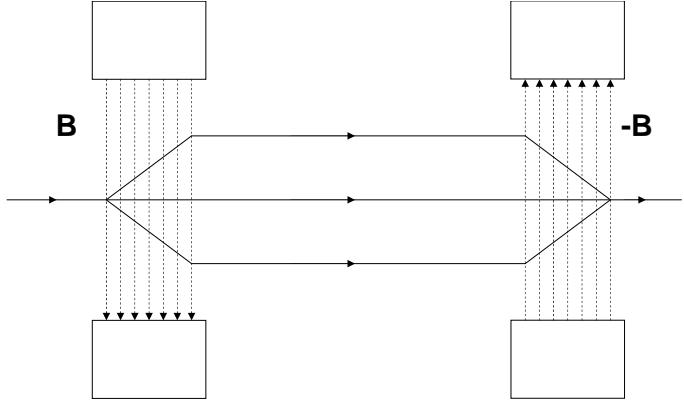}
\caption{Stern-Gerlach Apparatus} \label{Single Stern-Gerlach}
\end{center}
\end{figure}

In our gedankenversion of the Stern-Gerlach  experiment we imagine
spin 1 particles sent though a parallel, static, inhomogeneous
magnetic field $\textbf{B}$. By parallel we mean that vectors
$\textbf{B}(x,t)$ are parallel in the vicinity of the particle beam.
By static we mean that $\textbf{B}(x,t)$ is independent of t. We
shall also assume that the gradient of the $\textbf{B}$ field is
constant close to the beam. Now a particle with spin $\textbf{S}$
will have an effective magnetic moment in the $-\textbf{S}$
direction. Hence the particle will experience a force:
\begin{equation*}
{\bf F} \propto \nabla(\textbf{S}\cdot \textbf{B})
\end{equation*}
The force is proportional to the component of the spin
in the direction of the magnetic field and so
the beam will split into three
branches, corresponding to the three possible spin states. So for a
single particle a measurement ascertaining which branch holds the
particle constitutes a measurement of the component of the spin in
the direction of the magnetic field.
Note that we can thus measure the spin in any direction
other than along the path of the beam. However, if such a
measurement is not taken, the three branches can be coherently
recombined by application of the field $-\textbf{B}$ (see figure
\ref{Single Stern-Gerlach}).

The spin hilbert space $\cal{H}$ for a spin 1 particle is isomorphic
to $\mathbb{C}^{3}$. Let $S_{x}, ~S_{y},~S{z}$ denote the
observables corresponding to spin measurements in the x, y and z
directions. Then we can choose a basis of  $\cal{H}$,
$\{ \ket{0, z}, \ket{+1, z}, \ket{-1, z}\}$, in which  $S_{z}$
is diagonal. $S_{x}, ~S_{y},~S_{z}$ are then represented by the
standard $3\times 3$ spin matrices. A measurement of spin in the
$\textbf{B}$ direction corresponds to a basis in $\cal{H}$
consisting of the eigenvectors of
${\textbf{S.B}} =\text{B}_{x}S_{x}+\text{B}_{y}S_{y}+\text{B}_{z}S_{z}$.
The eigenvectors of spin in the $x$ and $y$ directions are
\begin{align}\label{eq:eigenvectors}
\ket{0,x} = &\frac{1}{\sqrt{2}}(\ket{+1,z} - \ket{-1,z})\, ,\\
\ket{+1,x} = &\frac{1}{{2}}(\ket{+1,z} + \sqrt{2} \ket{0,z} + \ket{-1,z})\, ,\\
\ket{-1,x} = &\frac{1}{{2}}(\ket{+1,z} - \sqrt{2} \ket{0,z} + \ket{-1,z})\, ,\\
\ket{0,y} = &\frac{1}{\sqrt{2}}(\ket{+1,z} + \ket{-1,z})\, ,\\
\ket{+1,y} = &\frac{1}{{2}}(\ket{+1,z} + i\sqrt{2} \ket{0,z} - \ket{-1,z})\, ,\\
\ket{-1,y} = &\frac{1}{{2}}(\ket{+1,z} - i\sqrt{2} \ket{0,z} - \ket{-1,z})\, .\label{eq:eigenvectors end}
\end{align}

\subsection{Using  Stern-Gerlach Apparatus to realise the Peres-Kochen-Specker setup}

Instead of spin we will be interested in spin squared,
{\it i.e.} $S_i^2$ in the $i$ direction (no sum on $i$).
There are now
two possible outcomes of a
measurement, $S_i^{2}=0$ (corresponding to $S_i=0$) and
$S_i^{2}=1$ (corresponding to $S_i=+1$ or $S_i=-1$). This can be
mirrored within the Stern-Gerlach framework by
lumping the two outer ($S_i=+1$ and $S_i=-1$)
beams together and labelling them
together as ``the red beam'' and labelling the middle
beam as ``the green beam''. This can be done ``mentally''
by simply ignoring the fine grained detail of which of
the outer beams the particle is in,
or ``physically'' by coherently recombining the two outer
beams into a single beam using a reversed Stern-Gerlach apparatus
(whilst keeping the middle beam separated by diverting
it out of the way).  Since in Anhomomorphic Logic we
may not be able to reason (about fine and coarse graining
for example) using classical rules,
it will be clearer to assume that we have set up a physical recombiner
so there really is a single ``red beam'' corresponding to
$S_i^2 = 1$. Let us further imagine appending the exact reverse of
this apparatus at the end which will
coherently recombine the red and the green beams into
a single beam again. We call this whole apparatus a
``spin-squared beam splitter and recombiner (bsr) in the
$i$ direction.''

For
a spin-1 particle, $S_i^2$ and $S_j^2$ commute if
the $i$ and $j$ directions are orthogonal, {\textit {i.e.}} the
squared $3\times 3$ spin representation
matrices $\sigma_{x}^{2},
~\sigma_{y}^{2},~\sigma_{z}^{2}$ commute.
In the standard Copenhagen
interpretation three (consecutive or simultaneous)
spin-squared measurements in mutually orthogonal directions will
necessarily result in one outcome of $0$ and two
outcomes of $1$.

This can be seen directly in terms of projectors onto the relevant
eigenspaces in the Hilbert space. Let $P_i^0 = \ket{0,i}\bra{0,i}$
and $P_i^1 = {\bf 1} - P_i^0$,
be projectors onto the spin-squared (in direction $i$)
eigenspaces corresponding to eigenvalues 0 and 1 respectively.
   Then, looking at equations \ref{eq:eigenvectors}-\ref{eq:eigenvectors end},
we have $P_i^0 P_j^0 = 0$ for any orthogonal pair of directions
$i$ and $j$, and for a basis $u_i,u_j,u_k$ we have $P_i^a P_j^b P_k^c = 0$ unless exactly one
of $a$, $b$ or $c$ equals 0.

To translate this into spacetime paths, we imagine
a spin-1 particle (in any spin state)
passing through a sequence of three
spin-squared bsr's, one in each of the three orthogonal
directions (no one of which coincides with the
direction of motion of the particle). The classical realist picture
is that the particle must pass through the green beam
in exactly one bsr and through the red beam in the other
two. We will see that the Peres-Kochen-Specker Theorem means that
this is untenable.

We want to translate colourings of the entire Peres set
$\{u_i\}$, not just one
basis, into spacetime paths. We imagine a
sequence of 33 spin-squared bsr's in the directions
$\{u_i\}$, so the first bsr will be
in direction $u_{1}$, the second in $u_2$, {\textit etc.}
The particle trajectories form the space $\Omega$
in this setup and each one
follows either the
red or green beam through each bsr in turn and
so every colouring can be realised
by an element of $\Omega$.
Strictly, there
are many particle trajectories in each beam, with slight
variations in positions, but we
will ignore this finer grained detail and assume that
$\Omega$ consists of the $2^{33}$ trajectories
distinguished only by which beam is passed through in
each bsr. This space is
then in one-to-one correspondence with the space of
colourings of the Peres Set and we identify the two in the
obvious way, noting that the particle paths contain additional
information, namely the choice of {\textit {order}} of the
Peres rays in the experimental set up, not present in the
original space of colourings.


Let the initial spin state of the particle be $\ket{\psi}$. Then a
decoherence functional, and hence a quantum
measure, on $\Omega$ can be defined as follows.
Let $\gamma$ be an element of $\Omega$, so $\gamma(u_i)$ is
a colour
for each Peres direction $u_i$.
As we saw above, colour ``green'' is identified with spin-squared
value zero and  colour ``red'' is identified with spin-squared
value one. We can construct a
``path state'' $\ket{\gamma}$ via
\begin{equation*}
\ket{\gamma} = P_{33}^{\gamma(u_{33})} \dots
P_2^{\gamma(u_2)} P_1^{\gamma(u_1)} \ket{\psi}
\end{equation*}
where $P_i^{green} \equiv P_i^{0}$ and $P_i^{red}\equiv P_i^{1}$
are the projection operators defined previously.
For each event $A\in \EA$ we can define an ``event state''
\begin{equation*}
\ket{A} = \sum_{\gamma \in A} \ket{\gamma}
\end{equation*}
and
the decoherence functional is then defined
by
\begin{equation*}
D(A;B) = \braket{A}{B}
 \,.
\end{equation*}

If the spin state of the particle is mixed the
decoherence functional is a convex combination
of such terms.

\begin{claim}
The quantum measure on $\Omega$ defined as above
values the PKS sets and their disjoint unions zero.
\end{claim}

\begin{proof}
Consider one of the PKS sets, $R_{\{u_i,u_j,u_k\}}$, say for a basis $\{u_i,u_j,u_k\}$.
Then the ``event state'' $\ket{R_{\{u_i,u_j,u_k\}}}$ is given by a
sum over ``path states'' $\ket{\gamma}$, one for each
$\gamma$ in $R_{\{u_i,u_j,u_k\}}$. This sum involves a sum over the
colouring of all the Peres rays which are not
$u_i$, $u_j$ or $u_k$ and so the projection operators for
all these rays sum to the identity and leave:
\begin{equation*}
\ket{R_{\{u_i,u_j,u_k\}}} = P_i^{0}P_j^{0}P_k^{0}\ket{\psi}
\end{equation*}
which equals zero for any state $\ket{\psi}$ because the
product of those projectors is zero. Similarly, the
event state for each of the PKS sets is zero because
it is a product of projectors acting on the initial
state and that product of projectors is zero.

An event which is a disjoint union of PKS events
has a corresponding event state
which is a sum of terms, one for each PKS event
in the union,
each of which is zero.

Hence result.
\end{proof}

For a given initial state and a choice of
ordering for the bsr's
we therefore have an explicit realisation of a
quantum measure in which the
PKS sets have measure zero. The statement of the Peres-Kochen-Specker
theorem in the context of this gedankenexperiment is that
every trajectory that a spin 1 particle can take
through the apparatus is in one of the
PKS precluded sets.

\subsection{Physical Reality}

Now consider the results of the previous section and their
application to a quantum measure of the form
constructed here (for some choice of linear ordering of the rays and a choice of
initial state). The preclusion rule that we wish to
impose requires that $|A|=0$ implies $\phi(A)=0$. The
co-event $\phi_M$ and its rotations constructed in the previous section
are preclusive on the PKS sets but they are not necessarily
preclusive on any other events of measure zero that might arise.
And there necessarily will be more sets of measure zero.

For example, if the choice of ordering of the Peres rays is
such that two orthogonal rays are consecutive, then any
element of $\Omega$ in which those two rays are coloured
green will be a (singleton) measure zero set because the
relevant projectors will be adjacent to each other and
they are orthogonal. Similar ``accidental'' measure
zero sets will arise whenever products of projectors that
are zero are either there from the beginning in the choice of ordering or
result from summing out intermediate projectors during coarse graining.
There may be more measure zero sets, depending
on $\ket{\psi}$, amongst the
``inhomogeneous'' sets {\textit i.e.} those that are not constructed by
summing over the sets of projectors for certain Peres rays.

We would like to use the co-event $\phi_M$ of the
previous section and analyse the consequences of considering
it as a potential reality. In order to make this analysis
fully meaningful, therefore, we need to find a measure
(an ordering of the rays and an initial state)
such that $\phi_M$ is preclusive. This would be accomplished
if the support of $\phi_M$, the set containing the two trajectories $\gamma_P$
and $\gamma_P'$, was not contained in any set of measure zero.
We believe that such a measure exists
but we are not able to give a proof,
the difficulty lying in the size of $\Omega$: $2^{33}$ colourings
means that the number of possible measure zero sets is
very large indeed.

We will, for the time being, assume that such a measure exists and
analyse $\phi_M$ accordingly. In other words we
assume that co-event $\phi_M$ is preclusive and so can
describe ``what actually happens'' at the microscopic level
in one run of the gedankenexperiment on a single particle.
In that case, what of the question posed at the end of
section 4?

Recall that in $\phi_M$ both the event ``ray 021 is green"
and the event ``ray 021 is red" are false. Translated into
spacetime terms we say that the particle is neither in the
red beam nor in the green beam of the 021 bsr.
The reason for this breakdown in the classical rule of
inference is that, of the two elements of the support of
$\phi_M$, $\gamma_P(021) =red$ and $\gamma_P'(021)=green$.
So neither redness nor greenness of ray 021
are common properties of the support of $\phi_M$ and so
neither is a real property of the world.

Understanding how this comes about at the abstract level
doesn't however remove the uneasy feeling that there is a
violation of known experimental facts here. Surely,
whenever we check to see whether the particle is in
the red beam or the green beam for any ray,
by putting detectors in both beams for example, we find it in one or the
other, not in neither.
This would only be a contradiction, however,
if such a measurement simply reveals, without
altering, the microscopic reality.
But, as is well known,
the presence of detectors alters the measure by destroying
coherence between beams
and so will change the possible co-events. The presence of
detectors in the 021 bsr will alter the measure so
that the $\phi_M$ co-event will no longer be preclusive and
no longer a possible reality.

It is tempting to call this ``contextuality'': the microscopic
reality depends on the experimental setup. Contextuality
always sets up a threat to (what in the literature is called) locality
and it is easy to see how that can occur: the ``context'' on
which real events over here depend could be the experimental setting of a
distant apparatus over there. In order to study this, we can
consider a system of two spin 1 particles in a spin zero
state so that their spin squared values are perfectly
correlated. These two particles are sent to distant
laboratories where they each pass through an
identical Peres-Kochen-Specker apparatus of 33 bsr's
as described above. This is under investigation.

Note that there is one case in which placing detectors
in both beams of the 021 bsr would not alter the measure and so {\textit{would}}
reveal the underlying microscopic reality without changing it.
That is if the 021 bsr is the last one in the sequence,
if 021 were the 33rd ray, $u_{33}$.
However, in this case, $\phi_M$
would not be preclusive for the following reason. $\gamma_P$
corresponds to a product of projectors (a ``class operator'')
and three of those projectors are onto $red$ for the three
rays in the basis $B_{11}$, one of which is $021$ and at the
very end of the product. $\gamma_P$ is therefore in a set of
measure zero which corresponds to summing over the projectors
(coarse graining over the colours)
for all the rays in between the three $B_{11}$ rays. The class
operator for
$\gamma_P'$, on the other hand, contains three projectors onto
$red$ for the basis rays in $B_7$ and so $\gamma_P'$ is
contained in a set of measure zero
that corresponds to summing over the projectors for
all the rays in between the three $B_7$ rays. These
two measure zero sets are disjoint, because $\gamma_P(u_{33}) =red$
and $\gamma_P(u_{33}) = green$ and the 33rd projector is
not summed over {\textit {because it is at the end and
not between any of the relevant projectors}}.
The disjoint union of these two measure zero sets is measure zero and
it contains both $\gamma_P$ and $\gamma_P'$, the support of $\phi_M$.
$\phi_M$ is, therefore, not a possible reality for this experimental
setup and there is no contradiction with experimental facts.

This result -- that no violations of classical rules of inference
can occur when restricting attention to events that occur at the
final time -- is a special case of a more general theorem
of Sorkin which will appear in a separate work.

\section{Discussion}

We have reviewed Sorkin's Multiplicative Schemes
for anhomomorphic logic in quantum measure theory.
We have shown  how to cast the Peres version of the Kochen-Specker
theorem into the framework of quantum measure theory
and described a Peres-Kochen-Specker gedankenexperiment
as a sequence of 33 beam-splitters and recombiners (bsl).
Certain subsets of paths of a spin 1 particle through this
apparatus have measure zero and are precluded and the Peres-Kochen-Specker
theorem says that the union of these subsets is the whole
space of paths, $\Omega$. Therefore no single path can be real.

We constructed explicit primitive co-events
in the Multiplicative Scheme which respect the PKS preclusions
however we do not know if there are quantum measures for
real experimental setups
for which these co-events respect {\it all} the preclusions.
We nevertheless used one of these explicit co-events to
begin an analysis of issues such as contextuality in
the scheme.

We present these results as an indication of the sorts of
questions that arise as we struggle towards an interpretation
of Quantum Measure Theory.
To investigate further, we need to study  examples of experimental
situations for much simpler ``quantum antimonies''
for which all the measure zero sets can be explicitly found and
all the primitive co-events calculated. This is being done
and an example due to  Hardy \cite{Hardy:1992, Hardy:1993} has been analysed by
Furey and Sorkin \cite{Furey:2007}.
We also need to work towards a measurement theory: any realistic interpretation
of quantum mechanics worth its salt must be able to reproduce the
predictions of Copenhagen quantum mechanics in the laboratory.

One motivation for pursuing
this line of research is the desire to restore objective reality
to physics, so that microscopic events can be
understood to be just as real as macroscopic ones.
The present authors are further motivated by the
expectation that
the correct interpretation of quantum mechanics will lead to
advances in {\it new} physics and specifically in quantum
gravity. It may, in particular, hold the explanation
of how -- Bell's inequalities notwithstanding --
physics can be both quantal and relativistically causal.
The reason that the framework of quantum measure theory
and anhomomorphic logic holds out this possibility is that it
incorporates real events that have no probability distribution on
them. It is possible then that such quantal events
could be the ``common causes''
of the Bell Inequality violating correlations. Such a
development could then lead to a
 formulation of a causal quantum dynamics for
spacetime itself, along the lines of the
formulation of causal classical stochastic dynamics \cite{Rideout:1999ub}.
Turning this around,
perhaps fruitfulness for quantum gravity
is the criterion whereby the correct interpretation
of quantum mechanics will eventually be recognised.

\section{Acknowledgments}
We thank Rafael Sorkin for help and discussions throughout
the course of this work. YG is supported by a PPARC
studentship. FD is supported in part by
the EC Marie Curie Research and Training Network,
Random Geometry and Random Matrices MRTN-CT-2004-005616
(ENRAGE) and by Royal Society International Joint Project 2006/R2.
FD thanks the Perimeter Institute for Theoretical Physics
for hospitality during the latter stages of this work.

\bibliography{../../../Bibliography/refs}
\bibliographystyle{../../../Bibliography/JHEP}
\end{document}